\documentclass[reprint,
superscriptaddress,
amsmath,amssymb,
aps,
floatfix,
longbibliography]{revtex4-1}
\setcitestyle{super}
\usepackage[colorlinks = true,
            linkcolor = blue,
            urlcolor  = blue,
            citecolor = blue,
            anchorcolor = blue,
            unicode]{hyperref}
\usepackage{graphicx}
\usepackage{amsmath}
\usepackage{amssymb}
\usepackage{amsfonts}
\usepackage{color}
\usepackage{xcolor}  
\usepackage{titlesec}

\titleformat{\section}
  {\normalfont} 
  {}{0pt}                      
  {}                           
\titleformat{\subsection}
  {\normalfont\bfseries} 
  {}{0pt}                      
  {}                           
\titlespacing*{\subsection}{0pt}{1.5ex}{4ex}
\makeatletter
\newcounter{ExtFig}

\newenvironment{ExtFigure}[1][htbp]{%
  \begin{figure*}[#1] 
    \let\c@figure\c@ExtFig 
}{%
  \end{figure*}
}
\makeatother
\begin{document}

\title{Electrically tunable moiré hybridization in scalable twist-with-a-spacer structures}
\newcommand{\equalcontrib}{\thanks{These authors contributed equally to this work.}}
\newcommand{\corresponding}{\thanks{Correspondence related to this work should be addressed to: alexey@nus.edu.sg; vladimir.falko@manchester.ac.uk; camilla.coletti@iit.it}}
\author{Ian Babich}\equalcontrib 
\affiliation{Department of Materials Science and Engineering, National University of Singapore, Singapore, Singapore}
\affiliation{Institute for Functional Intelligent Materials,
National University of Singapore, Singapore, Singapore}
\author{Daria Belotcerkovtceva}\equalcontrib  
\affiliation{Institute for Functional Intelligent Materials,
National University of Singapore, Singapore, Singapore}

\author{Benjamin Fuller}\equalcontrib  
\affiliation{Department of Physics and Astronomy, University of Manchester, Manchester, UK}
\affiliation{National Graphene Institute, University of Manchester, Manchester, UK}

\author{Ilya Begichev}  
\affiliation{Department of Materials Science and Engineering, National University of Singapore, Singapore, Singapore}
\affiliation{Institute for Functional Intelligent Materials,
National University of Singapore, Singapore, Singapore}

\author{Nansi Zhou}  
\affiliation{Department of Materials Science and Engineering, National University of Singapore, Singapore, Singapore}
\affiliation{Institute for Functional Intelligent Materials,
National University of Singapore, Singapore, Singapore}

\author{Alex Boschi}
\affiliation{Center for Nanotechnology Innovation @NEST, Istituto Italiano di Tecnologia, 56127 Pisa, Italy}

\author{Vaidotas Miseikis}
\affiliation{Center for Nanotechnology Innovation @NEST, Istituto Italiano di Tecnologia, 56127 Pisa, Italy}

\author{Gianluigi Baiardi}
\affiliation{Center for Nanotechnology Innovation @NEST, Istituto Italiano di Tecnologia, 56127 Pisa, Italy}

\author{Andrei Kudriashov}  
\affiliation{Department of Materials Science and Engineering, National University of Singapore, Singapore, Singapore}

\affiliation{Institute for Functional Intelligent Materials,
National University of Singapore, Singapore, Singapore}

\author{Konstantin S. Novoselov}  
\affiliation{Department of Materials Science and Engineering, National University of Singapore, Singapore, Singapore}
\affiliation{Institute for Functional Intelligent Materials,
National University of Singapore, Singapore, Singapore}

\author{Stiven Forti}
\affiliation{Center for Nanotechnology Innovation @NEST, Istituto Italiano di Tecnologia, 56127 Pisa, Italy}

\author{Antonio Rossi}
\affiliation{Center for Nanotechnology Innovation @NEST, Istituto Italiano di Tecnologia, 56127 Pisa, Italy}

\author{Fabio Beltram}
\affiliation{NEST Laboratory, Scuola Normale Superiore, 56127 Pisa, Italy}

\author{Kenji Watanabe}  
\affiliation{Research Centre for Electronic and Optical Materials, National Institute for Material Science, Tsukuba, Japan}

\author{Takashi Taniguchi}  
\affiliation{Research Centre for Materials Nanoarchitectonics, National Institute for Material Science, Tsukuba, Japan}

\author{Sergio Pezzini}
\affiliation{Istituto Nanoscienze–CNR, NEST-Scuola Normale Superiore, 56127 Pisa, Italy}

\author{Camilla Coletti} \corresponding
\affiliation{Center for Nanotechnology Innovation @NEST, Istituto Italiano di Tecnologia, 56127 Pisa, Italy}

\author{Vladimir Fal'ko} \corresponding 
\affiliation{Department of Physics and Astronomy, University of Manchester, Manchester, UK}
\affiliation{National Graphene Institute, University of Manchester, Manchester, UK}

\author{Alexey I. Berdyugin} \corresponding 
\affiliation{Department of Materials Science and Engineering, National University of Singapore, Singapore, Singapore}
\affiliation{Department of Physics, National University of Singapore, Singapore, Singapore}
\affiliation{Institute for Functional Intelligent Materials,
National University of Singapore, Singapore, Singapore}

\begin{abstract}
Moiré superlattices emerge from the interplay between interlayer hybridization and long-wavelength periodicity, both set by the twist angle between layers in direct contact. Here we show that moiré minibands and pronounced Brown-Zak oscillations persist even when twisted graphene layers are separated by an atomically thin spacer, such as a misaligned graphene monolayer. This twist-with-a-spacer architecture reduces interlayer coupling, such that moiré hybridization can be tuned with a modest displacement field without changing moiré periodicity, allowing us to disentangle the roles of those parameters in miniband formation. Importantly, the spacer suppresses atomic relaxation, enabling highly uniform, rigid superlattices at twist angles down to $0.1^{\circ}$. This robustness allows twist-with-a-spacer structures to be grown directly by scalable chemical vapour deposition (CVD): devices fabricated from CVD-grown structures exhibit the same moiré physics phenomena as those assembled manually. Our work provides insights into the physics of moiré bandstructure formation and opens a route towards scalable moiré systems.
\end{abstract}

\maketitle
\subsection*{Introduction} \vspace{-1.5em}
    Controlling the coupling between several two-dimensional electronic systems by tuning their mutual orientation and separation is a powerful method for engineering novel electronic states and quantum phenomena inaccessible in individual layers \cite{andrei2020graphene, lau_reproducibility_2022, mak_semiconductor_2022, li2010observation, bistritzer2011moire, cao2016superlattice, kim2017tunable, cao2018unconventional, lu2019superconductors, uri2020mapping, willett1987observation, rojo1999electron, eisenstein2004bose, ponomarenko2011tunable, kim2011coulomb, gorbachev2012strong, narozhny2016coulomb, liu2017quantum, li2017excitonic, britnell2012electron, amet2012tunneling, britnell2012field, georgiou2013vertical, britnell2013resonant, mishchenko2014twist}. In particular, recent research has focused on small-angle twisted bilayers, where strong interlayer hybridization, combined with an additional long-wavelength periodicity, leads to the formation of moiré minibands \cite{andrei2020graphene, lau_reproducibility_2022, mak_semiconductor_2022, li2010observation, bistritzer2011moire, cao2016superlattice, kim2017tunable, cao2018unconventional, lu2019superconductors, uri2020mapping}. In this respect, the interlayer distance serves as an important control parameter that simultaneously affects two fundamental energy scales: the amplitude of electronic tunneling across the interface and the strength of Coulomb interactions between carriers in different layers. At large separations ($\sim$2–10 nm), interlayer tunneling is fully suppressed, and the layers interact only through long-range forces \cite{willett1987observation, rojo1999electron, eisenstein2004bose, ponomarenko2011tunable, kim2011coulomb, gorbachev2012strong, narozhny2016coulomb, liu2017quantum, li2017excitonic}. Reducing the spacing to the nanometer range ($\sim$1–2 nm) enables measurable tunneling currents \cite{britnell2012electron, amet2012tunneling, britnell2012field, georgiou2013vertical, britnell2013resonant, mishchenko2014twist}, though insufficient for any noticeable reconstruction of a bandstructure. Hence, it is generally assumed that the formation of moiré minibands occurs only when the two crystals are in direct atomic contact, providing strong interlayer tunneling \cite{andrei2020graphene, lau_reproducibility_2022, mak_semiconductor_2022, li2010observation, bistritzer2011moire, cao2016superlattice, kim2017tunable, cao2018unconventional, lu2019superconductors, uri2020mapping}. 
    
    Nevertheless, in the limit when two twisted layers are separated by a single-layer spacer -- despite being weaker than in conventional twisted heterostructures -- the interlayer tunnelling potentially could still reconstruct the overall electronic bandstructure. This expectation is supported by earlier works on twisted MoSe$_2$ with a monolayer insulating spacer \cite{shimazaki2020strongly, schwartz2021electrically}, which demonstrated hybridization of inter- and intralayer excitons, as well as by the known role of next-layer hopping in graphite, where it affects band dispersions at low energies \cite{mcclure_band_1957, haering1958band, dresselhaus_intercalation_2002, zhang2010band}. However, despite an apparent residual tunnelling contribution naturally expected in this regime, so far, there has been neither direct experimental evidence that such weak tunnelling would induce notable moiré effects in twisted graphene layers separated by an incommensurate spacer, nor a detailed theoretical analysis of moiré minibands properties in this limit. Here, we demonstrate that tunnelling between twisted graphene layers separated by an atomically thin spacer remains sufficiently strong to reconstruct their individual band structures, giving rise to pronounced moiré effects in twist-with-a-spacer structures fabricated using very distinct approaches. 
    
\begin{figure*}
\begin{center}
	\includegraphics[width=\linewidth]{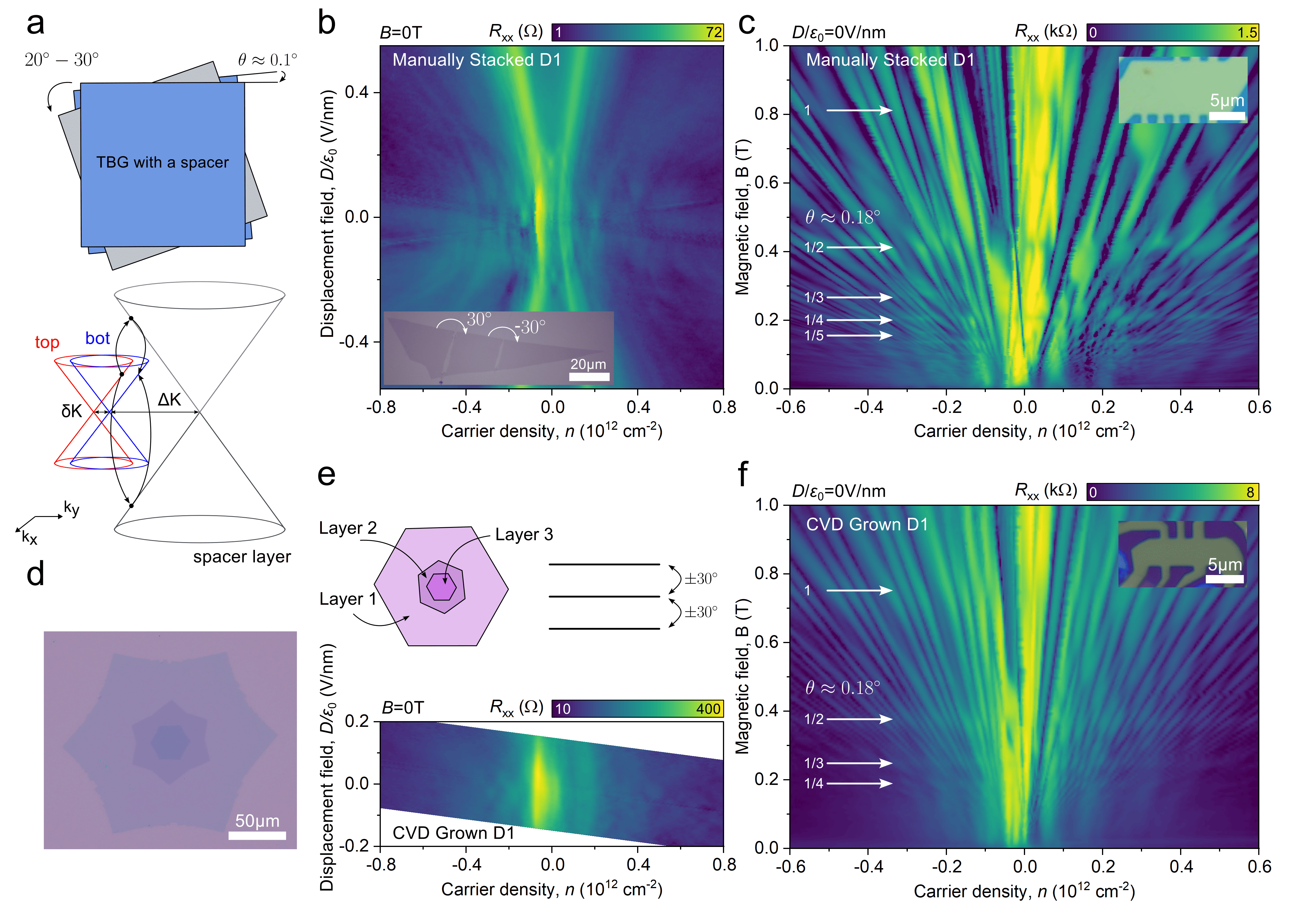}
    \caption{\textbf{Detecting coupling between graphene layers twisted through a monolayer spacer.}
    \textbf{a}, Schematic illustration of the twist-with-a-spacer structures used in this work. Top: TBG structure with a graphene spacer. The outer graphene layers are twisted by a small angle relative to each other and decoupled from the middle layer by a large twist angle. Bottom: the corresponding configuration in momentum space. $\delta K$ and $\Delta K$ indicate the momentum-space separations between the Dirac points arising from the small and large twist angles, respectively. Arrows indicate two hopping pathways between the top and bottom layers mediated by either the conduction or valence band of the graphene spacer.
    \textbf{b}, Longitudinal resistance as a function $n$ and $D$ for a device fabricated from manually stacked graphene layers (Manually Stacked D1). Inset: optical image of a representative exfoliated graphene flake cut into three parts, together with the sequence of twist angles used to assemble the stack.
    \textbf{c}, Landau fan diagram measured for Manually Stacked D1 at $D=0$. White horizontal arrows point to Brown-Zak oscillations, and labels indicate flux through a moiré unit cell normalized by the non-superconducting flux quantum. 
    \textbf{d}, Optical image of CVD-grown twisted graphene layers, in which the top and bottom layers are aligned, and the middle layer is rotated by approximately $30^\circ$ relative to them.
    \textbf{e}, Top: schematic illustrations of the crystal orientations (left) and relative twist angles (right) of the CVD-grown stack shown in (d). Bottom: resistivity as a function of $n$ and $D$ for a device assembled from the CVD-grown twisted layers (CVD-grown D1). Leakage through the top-gate dielectric in this device began at a relatively low gate voltage, limiting the accessible range of $D$.
    \textbf{f}, Landau fan diagram measured for CVD Grown D1 at $D=0$. As in panel (c), horizontal arrows indicate Brown--Zak oscillations, with labels denoting the magnetic flux through a moiré unit cell normalized. Inset: optical image of the device. All measurements were performed at $T$=2K.}
	\label{fig1}
\end{center}
\end{figure*}

    First, to explore this limit, we fabricated twisted bilayer graphene (TBG) devices based on exfoliated graphene layers with two graphene layers rotated by a small angle and separated by an additional graphene sheet that serves as a spacer. The twist geometry is schematically illustrated in Fig.~\ref{fig1}a (top panel), and the typical graphene flake cut into several parts used in our structures is shown in the inset of Fig.~\ref{fig1}b . The middle layer is effectively electronically decoupled from the outer layers by large twist angles, a well-established approach to suppress interlayer hybridization \cite{sanchez2017helical, rickhaus2020electronic, Slizovskiy2021, babich2025milli}. The effective coupling between the top and bottom graphene layers across the spacer is expected to be an order of magnitude weaker than between directly stacked twisted bilayers. To enhance the impact of the residual coupling, we intentionally align the outer graphene layers to a near-zero relative angle ($\theta \approx 0.1^{\circ}$). The resulting heterostructures are fully encapsulated in hexagonal boron nitride (hBN), with top and bottom graphite gates on both sides used for the independent control of the total carrier density, $n$, and vertical displacement field, $D$. 

    In addition to the devices assembled from exfoliated and manually twisted graphene, we studied graphene structures grown by chemical vapor deposition (CVD) using a protocol that produces 30$^\circ$-twisted interfaces between consecutive layers \cite{yan2014large, pezzini202030, cho2020specific}. Briefly, during graphene growth on electropolished Cu, hexagonal monolayer crystals nucleate, and successive layers grow concentrically from the same nucleation center, sharing a common core. The orientation of each new layer is established at nucleation through the alignment of its straight zigzag or armchair edges with atomic step edges on the Cu(111) surface. This interaction strongly favours either complete alignment between consecutively grown layers (a twist angle of $0^{\circ}$, corresponding to Bernal-stacked bilayer graphene) or a $30^{\circ}$ configuration, in which the zigzag edge of one layer aligns with the armchair edge of the other \cite{yan2014large, pezzini202030, cho2020specific}. When a third layer is grown, it can be either aligned with the second layer or rotated by $30^{\circ}$ relative to it. One resulting configuration has nearly aligned top and bottom graphene layers separated by a $\sim30^{\circ}$-misaligned middle layer \cite{yan2014large}, which acts as an atomically thin spacer, analogous to that in the exfoliated devices discussed above. After growth, these structures were identified by optical microscopy, as shown in Fig.~\ref{fig1}d,e. Once a suitable CVD-grown trilayer crystal is identified, we use standard dry-transfer and nanofabrication procedures to fabricate an hBN-encapsulated dual-graphite-gated device. 

    In total, we have characterized six Hall bars: three fabricated from exfoliated and manually twisted graphene layers, and three from CVD-grown trilayers. All devices exhibit qualitatively similar magneto-transport behavior. Further fabrication details and device parameters are provided in the Supplementary Information Note 1. 

\subsection*{Results} \vspace{-1.5em}

\subsection*{Detecting coupling of twisted graphene layers through the spacer} \vspace{-1.5em}

    First, we characterize our devices made of exfoliated graphene layers. To this end, we first measured the longitudinal resistance $R_{\mathrm{xx}}$ as a function of carrier density and displacement field, Fig.~\ref{fig1}b. At $D=0$, several resistance peaks emerge near zero doping. When a displacement field is applied, most of these features disappear, while two dominant peaks persist and shift only weakly with increasing $D$. This is in a big contrast to the typical behavior observed in devices with decoupled graphene layers, which show only a single peak at $D=0$ that splits under increased displacement field into charge neutrality points of the individual layers \cite{babich2025milli}. The signatures of interlayer coupling are also prominent in the Landau-fan diagram measurements at $D=0$ Fig.~\ref{fig1}c. In this figure, multiple fans converge to distinct zero-field offsets, revealing the features typically related to moiré minibands edges. Clear Brown–Zak oscillations \cite{krishna2017high, krishna2018high} also appear as density-independent lines traceable across this map, with periodicity in $1/B$ corresponding to the twist angle $\theta \approx 0.18^{\circ}$. More examples of fan diagrams and Brown-Zak oscillations measured in other manually stacked devices are shown in Extended Data Figs.~\ref{si4},\ref{si3}. 

    Devices fabricated from CVD-grown trilayers were characterised using the same approach. These devices show the same signatures of interlayer coupling: two well-resolved peaks at zero magnetic field Fig.~\ref{fig1}e, together with clear Brown-Zak oscillations superimposed over the fan diagram displayed in Fig.~\ref{fig1}f. The twist angle extracted from those oscillations is $\theta \approx 0.18^{\circ}$, coincidentally close to that of the exfoliated device. Further data from this device are shown in Extended Data Fig.~\ref{CVD_data}. In addition, a similar as-grown CVD crystal was characterized using angle-resolved photoemission spectroscopy (ARPES), which was used to confirm the 30$^\circ$ rotation between the consecutive layers. An ARPES spectrum in Extended Data Fig.~\ref{CVD_data}a presents a K$_1$-$\Gamma$ line-cut for the top layer, whose Dirac cone dominates in the ARPES intensity and features a peculiar anticrossing at $\approx$-1.5~eV, identified earlier to be the result of a hybridization of K$_1$ and K'$_1$ cones induced by the umklapp scattering from the honeycomb lattice of the middle layer \cite{yao2018quasicrystalline}. Unfortunately, the spectral resolution of ARPES is insufficient to observe low-energy details caused by an ultra-long-period moiré superlattice, leaving electron transport as a unique tool for the characterization of such twist-with-a-spacer structures. 
    

\subsection*{Probing moiré physics in TBG-with-a-spacer} \vspace{-1.5em}

\begin{figure*}
    \begin{center}
	\includegraphics[width=\linewidth]{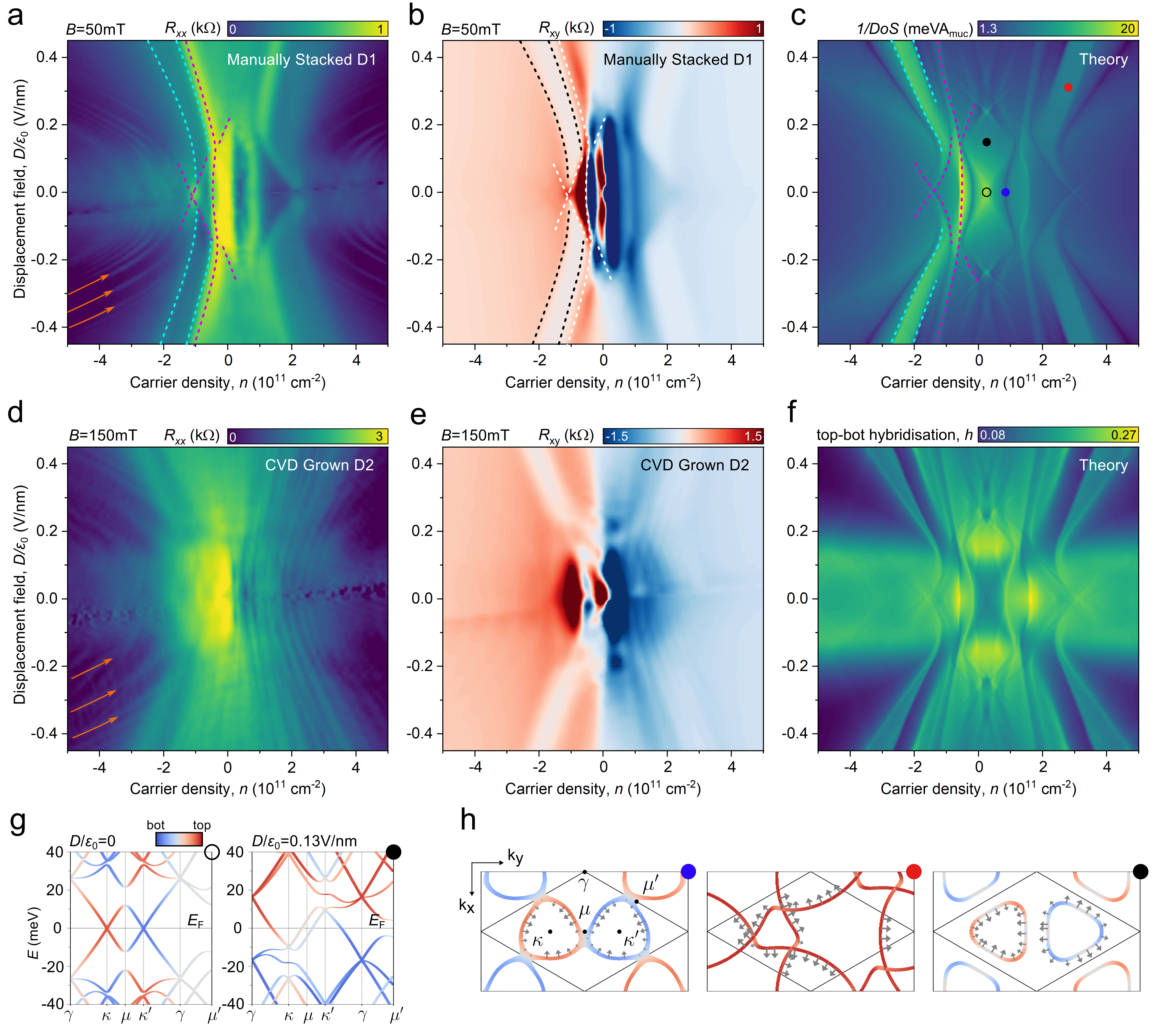}
    \caption{\textbf{Probing the bandstructure of TBG with a spacer.} 
    \textbf{a–b}, Longitudinal and Hall resistance of the Manually Stacked D1 measured at 50 mT as a function of $n$ and $D$. Measurements were performed at 2 K. In panels (a) and (b), dashed lines highlight features originating from two distinct mechanisms: direct interlayer hybridization, indicated by cyan and black lines, and umklapp tunneling, shown by magenta and white lines. See main text for further details. Orange arrows in the bottom left corner of (a) indicate Landau quantization of one of the layers.
    \textbf{c}, Inverse density of states (1/DoS) at the Fermi energy of TBG with a spacer as a function of $n$ and $D$. Cyan and magenta dashed lines show features originating from pure layer hybridization and umklapp scattering, respectively. Colored circles indicate ($n,D$) values at which Fermi contours and spectra shown in panels (g,h) were calculated. 
    \textbf{d–e}, Longitudinal and Hall resistance of the CVD Grown D2 measured at 150~mT as a function of $n$ and $D$. Measurements were performed at 2 K. Orange arrows in the bottom-left corner of (d) indicate Landau quantization of one of the layers.
    \textbf{f}, Layer hybridization of TBG with a spacer as a function of $n$ and $D$. See main text and Supplementary Note 5 for further details. 
    \textbf{g}, Bandstructures of the TBG device with a spacer calculated for the twist angle $\theta = 0.18^{\circ}$ for several values of $n$ and $D$, corresponding to the colored circles shown in panel (c). See Supplementary Note 5 and Supplementary Table 2 for calculation details. The color of the bands shows the fraction of the carriers shared by each layer according to the scale bar; e.g., red (blue) bands are strongly top (bottom) layer polarized. 
    \textbf{h}, Fermi contours of TBG with a spacer calculated for different $n$ and $D$ highlighted with colored circles in panel (c). Grey arrows indicate the local velocity of the band structure and thus the sign of the effective mass: arrows pointing inside the Fermi contour correspond to an electron-like Fermi surface, whereas arrows pointing outside indicate a hole-like Fermi surface. The colors reflect the layer polarization, similar to panel (g).
    }
	\label{fig2}
\end{center}
\end{figure*}

    In Fig.~\ref{fig2}a,b and Fig.~\ref{fig2}d,e  we show  $R_{\text{xx}}$ and $R_{\text{xy}}$ as functions of $D$ and $n$, measured at small magnetic fields in manually stacked sample D1 and CVD-grown sample D2, respectively (see Extended Data Figs.~\ref{si4},\ref{CVD_data} for further examples). The magnetic field amplifies the structures already visible at $B=0$, revealing a central 'rhombus' on the $n-D$ map featuring multiple resistance peaks and Hall-sign reversals. We also note persistent stripe-like features that shift with increasing $D$, along which $R_{\text{xy}}$ is suppressed (marked by black dashed lines in Fig.~\ref{fig2}b). In the corners of the $R_{\text{xx}}$ maps, additional peaks emerge, which we attribute to Landau levels (LLs) of the individual outer graphene layers, similar to those observed in decoupled twisted graphene devices \cite{babich2025milli}. These LL features abruptly weaken upon reaching a horizontal belt-like feature around $D=0$ with slightly increased resistance, which we attribute to the enhanced coupling between the outer graphene layers through a monolayer spacer (see details below). The close agreement between the transport maps of the manually stacked and CVD-grown devices demonstrates that both exhibit the same characteristic signatures of the interlayer coupling. In the CVD-grown device 2, Fig.~\ref{fig2}d,e, an additional set of nearly vertical lines is visible, likely originating from the electronically decoupled middle layer, as reported previously \cite{babich2025milli}. 

    To gain further insight into the observed magneto-transport features, we model the bandstructure of a TBG with a monolayer spacer. For this, we use a 6×6 $k\cdot p$-theory Hamiltonian incorporating twisted interfaces and the full Slonczewski–Weiss–McClure (SWMcC) parametrization of the interlayer coupling \cite{garcia-ruiz2021full}. An effective low-energy model for the outer layers was then constructed, eliminating the middle layer states using perturbation theory, where we keep terms of zeroth order in the valley momentum (counted from K$_{1,3}$-points of the outer layers). This effective interlayer coupling arises from two terms. The first is direct next-nearest-layer hopping, as in bulk Bernal-stacked graphite, but modified for twisted structures \cite{garcia-ruiz2021full}. That is, in Bernal graphite, the fixed stacking geometry distinguishes between the $\gamma_2$ and $\gamma_5$ next-nearest-layer hopping parameters, which correspond to inequivalent atomic positions in the intermediate layer \cite{dresselhaus_intercalation_2002}. Here, the twisted spacer layer removes this difference, so we estimate the direct tunneling contribution as the average of these parameters, as $\frac12 (\gamma_2+\gamma_5)$. The second contribution comes from virtual transitions through the spacer’s conduction and valence band states as illustrated in Fig.~\ref{fig1}a (bottom panel). As the contributions from the virtually occupied conduction and valence band states interfere destructively, the overall tunneling amplitude between the outer layers is sensitive to the electron-hole asymmetry in the multilayer Hamiltonian; it would be zero if studied in the minimal interlayer coupling model based on only $\gamma_1$-hopping among the SWMcC parameters, but is non-zero after we take into account $\gamma_3$ and $\gamma_4$ hopping processes \cite{dresselhaus_intercalation_2002}. 
    
    The resulting effective hopping parameter coupling the outer layers is estimated (see Supplementary Note 2) as 
\begin{equation}
    \label{3}
    \tilde{\gamma} = \frac{1}{3} \left(
    -\frac{2\gamma_{1}(\gamma_{3}-\gamma_{4})}{3\gamma_{0}}
    +
    \frac{\gamma_{2}/2+\gamma_{5}/2}{2}
    \right)
    \approx -5 \text{ meV},
\end{equation}
    where $\gamma_0$ is the in-plane nearest-neighbour hopping parameter \cite{dresselhaus_intercalation_2002}, and tunneling process through the virtual intermediate conduction/valence band states in the spacer layer is about 4 times larger than the direct tunneling contribution. This hopping parameter scales the tunneling matrix elements that couple A/B sublattices of the bottom/top graphene layers,
\begin{equation}
    \label{eq:t}
    \mathcal{T}_{\textbf{k}',\textbf{k}} = 
    \tilde{\gamma}
    \sum_{j=0}^2
\begin{pmatrix}
    1 & e^{i\xi \frac{2\pi}{3} j} \\ 
    e^{-i\xi \frac{2\pi}{3} j} & 1
\end{pmatrix}
    \delta_{\textbf{k}'\!,\, \textbf{k}
    +\delta\textbf{K}^{(0)}_{\xi}-\delta\textbf{K}^{(j)}_{\xi}},
\end{equation}
    where $\xi=\pm1$ is the valley index and $\delta\textbf{K}^{(j)}_{\xi} = \theta\hat{\textbf{z}}\times\textbf{K}^{(j)}_{\xi}$ is the momentum mismatch between $\textbf{K}^{(j)}_{\xi}=\xi K\left(\cos \frac{2\pi j}{3}, -\sin \frac{2\pi j}{3}\right)$ points of the outer layers as a result of the twist $\theta$. 
    This interlayer coupling model is applicable to general configurations of twist angles as long as the middle layer is sufficiently decoupled, that is, when $\delta K \ll \Delta K$ and $\hbar v \Delta K \gg \gamma_1/3$, where $\Delta K$ is the momentum shift of the spacer layer relative to the outer layers, and $v$ is the Dirac velocity of electrons in graphene.

    A representative band structure for $\theta=0.18^{\circ}$ at $D/\varepsilon_{0}=0$ is shown in Fig.~\ref{fig2}g. Despite the small twist angle, the Dirac cones of the individual layers remain essentially unchanged, in contrast to the conventional TBG spectra. This is because the effective coupling $\tilde{\gamma}$ of TBG with a monolayer spacer is 25 times smaller than the coupling of the directly stacked twisted layers \cite{bistritzer2011moire}. Correspondingly, the effective “magic angle" would be reduced in proportion, shifting towards $\theta_{\text{Magic}} \approx 0.04^{\circ}$, as shown in Extended Data Fig.~\ref{si_theory}. Nevertheless, the Fermi surfaces of the outer layers hybridize at energies $\sim$10 meV, which is consistent with the resistance features and Hall-sign reversals visible at $D/\varepsilon_{0}=0$ in Fig.~\ref{fig2}a,b,d,e.

    To further check the correspondence between the experiment and theory, we computed the density of states (DoS) at the Fermi energy as a function of $n$ and $D$ for the entire experimentally explored range. For this, we have implemented a self-consistent electrostatic model that accounts for the charge transfer between the layers, including the spacer layer (see in Supplementary Note 3). Since resistance scales inversely with DoS, we compare $1/\mathrm{DoS}$, Fig.~\ref{fig2}c, with the $R_{\text{xx}}$ maps, Fig.~\ref{fig2}a,d. Moreover, by modeling conductivity in a small magnetic field, we also calculated the Hall resistance (see Extended Data Fig.~\ref{si_theory}). We note that all maps exhibit an intrinsic asymmetry because the spacer layer neutrality point is shifted in energy due to the moiré interactions in the effective model, which offsets the fillings in the outer layers and spacer layer at each CNP (see Supplementary Note 3). We also note that the position of the rhombus-shaped region and those of the other structures in Fig.~\ref{fig2}c are sensitive to the twist angle, allowing us to use these modeling results as an additional cross-check for twist-angle identification based on Brown–Zak oscillations. 
    
    The simulated DoS and Hall maps reproduce the main experimentally observed features well, including the rhombus-shaped region with vertical lines, multiple Hall-sign reversals, and the stripes at large $D$. To identify the origin of these features, we examined Fermi contours, cut through the miniband spectra, as shown in Fig.~\ref{fig2}h. At $D/\varepsilon_0 = 0$, increasing $n$ drives a crossing of the top- and bottom-layer Fermi surfaces (empty blue circle panel in Fig.~\ref{fig2}h), producing a Lifshitz transition that coincides with resistance peaks and Hall-sign changes in measured $n$-$D$ maps. Also, within the stripe-like regions (marked by a red circle in Fig.~\ref{fig2}c), the bands become strongly layer-polarized, confining the charge carriers predominantly to one layer. In this regime, the system enters a metallic state characterized by relatively large Fermi contours (Fig.~\ref{fig2}h), resulting in the experimentally observed suppression of both the longitudinal resistivity and Hall coefficient. 

        \begin{figure*}
\begin{center}
	\includegraphics[width=\linewidth]{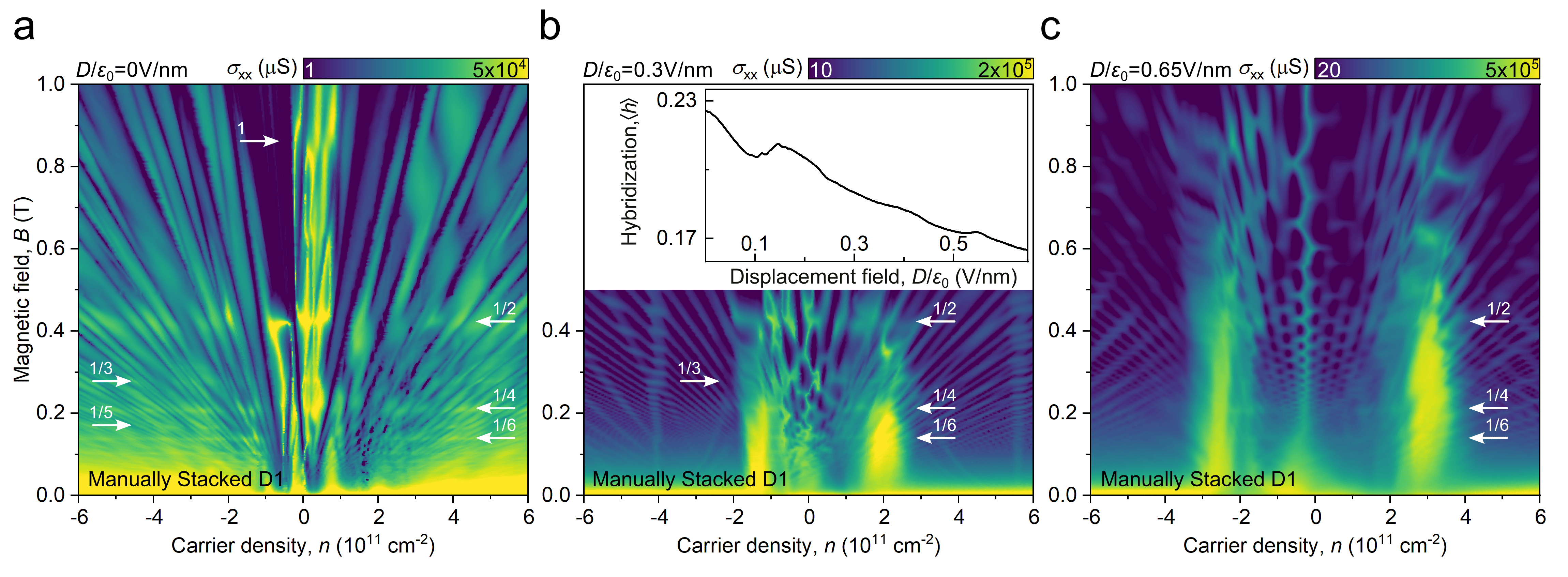}
	\caption{\textbf{Tunability of Brown-Zak oscillations}.
    \textbf{a}, $\sigma_{\text{xx}}$ as a function of carrier density $n$ and magnetic field $B$ at zero displacement field. Black arrows point at Brown-Zak oscillations 
    \textbf{b}, The same as (a), but $D/\varepsilon_0=0.3\,$V/nm. Note that Brown-Zak oscillations under applied $D$ are contained between the two bright peaks that are also seen in double-gate maps in Fig.~\ref{fig2}a, b as resistivity stripes and Hall-effect reset stripes. The inset shows hybridization as a function of displacement field averaged in the $n=\pm2\times 10^{11}\,\text{cm}^{-2}$ window. 
    \textbf{c}, Same as (a) and (b), but $D/\varepsilon_0=0.65\,$V/nm. The honeycomb-shaped patterns in panels (b) and (c) at high magnetic fields originate from crossings between the Landau levels of the top and bottom layers, similar to those observed in decoupled graphene layers \cite{babich2025milli}.
    All measurements were performed on Manually Stacked D1 at 2K.} 
	\label{fig3}
\end{center}
    \end{figure*}
    
    Finally, we analyzed the evolution of Fermi contours upon the variation of a displacement field, $D$. At zero $n$, the applied $D$ introduces an energy offset between the layers, driving the system into a semimetallic state with electrons in one layer and holes in the other. Moreover, as the group-velocity vectors of states in different layers coincide, electrons become effectively shared between both layers over the whole $\kappa-\kappa'$ branch of the hybridized dispersion. Increasing $D$ enlarges the corresponding Fermi pockets until a resonant tunneling condition is reached, at $D/\varepsilon_0 \approx 0.13\,$V/nm, which is marked by a black circle in Fig.~\ref{fig2}c, with corresponding dispersions and Fermi contours shown in Fig.~\ref{fig2}g and Fig.~\ref{fig2}h (panel with black circle) \cite{britnell2013resonant}. At this point, the layer polarization is strongly suppressed over large portions of the Fermi surface. More detailed evolutions of Fermi contours (for $D=0$ with changing $n$, and for $n=0$ with changing $D$) are shown in Supplementary Videos 1 and 2.

    Moreover, we also identify what features of the bandstructure are determined by momentum-conserved tunneling between the layers (like in G/hBN/G tunneling devices in Ref. \cite{mishchenko2014twist}), and the features truly originating from moir{\'e} superlattice minibands related to umklapp processes (tunneling accompanied by a momentum boost by moiré Bragg vector). To this end, we compare the full calculation to a toy model corresponding to two tunneling-coupled Dirac systems offset in momentum space, which we obtained by keeping only the $j=0$ term in the tunneling matrix in Eq.~\ref{eq:t}. Extended Data Fig.~\ref{si_theory} illustrates the calculated 1/DoS maps in which the coupling effects of the outer layers are included step by step. We start from a purely electrostatic model, Extended Data Fig.~\ref{si_theory}d; then the results of the toy model, Extended Data Fig.~\ref{si_theory}e, and, then the full moiré superlattice model, Extended Data Fig.~\ref{si_theory}f. For example, we identify the areas marked on the map in Fig. ~\ref{fig2}c by three filled circles caused by the interlayer tunneling without umklapp.  When the complete set of couplings Eq.~\ref{eq:t} was included, progressively more features emerged in the DoS, in particular, those marked by dashed magenta lines. Notably, many of these reproduce the details of the experimentally observed maps in Fig.~\ref{fig2}a,b,d,e and Extended Data Fig.~\ref{si4}d,e. 
    
    \begin{figure*}
\begin{center}
	\includegraphics[width=\linewidth]{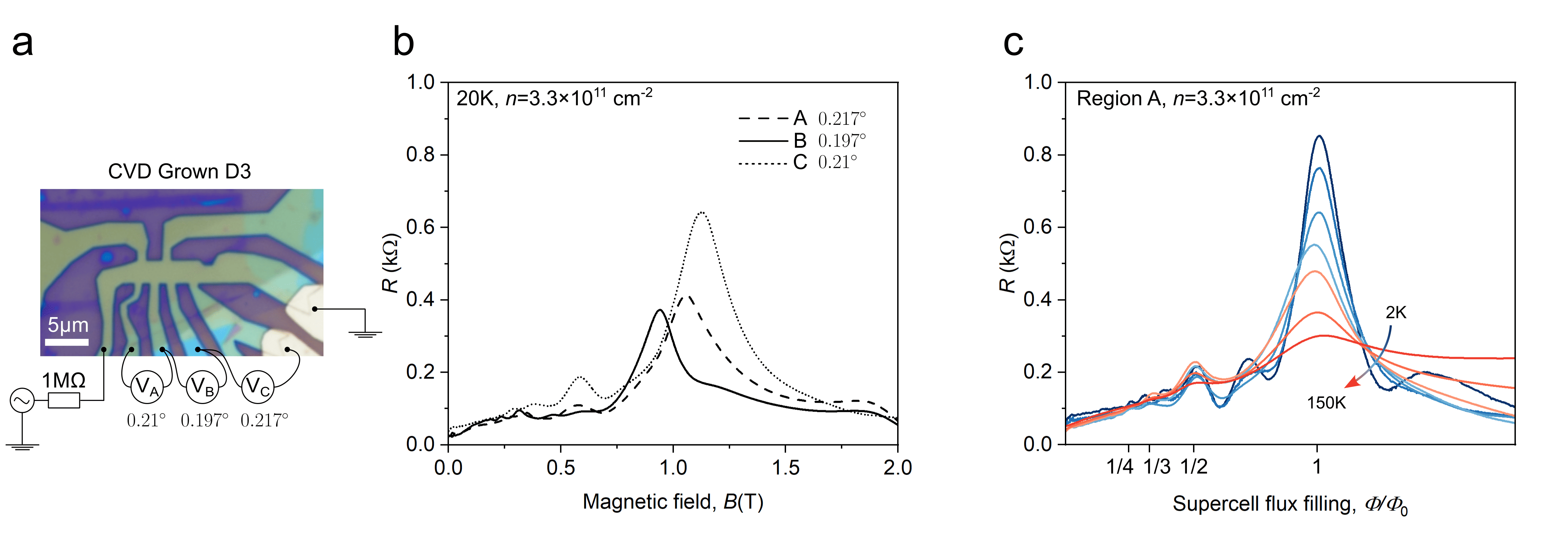}
    \caption{\textbf{Characterization of twist-angle inhomogeneity in TBG-with-a-spacer.}
    \textbf{a}, Optical image of CVD-grown D3, with a schematic illustration of the measurement geometry used to characterize spatial variations in the twist angle.
    \textbf{b}, Brown-Zak oscillations measured at T=20 K and fixed carrier density $n$ in different regions of CVD Grown D3, as indicated in (a).
    \textbf{c}, Brown-Zak oscillations measured at different temperatures in region A indicated in (a).}
\label{fig4}
\end{center}
    \end{figure*}
    
    \subsection*{Controllable hybridization and layer polarization} \vspace{-1.5em}
    
    We find that a strong and gate-tunable hybridization between the twisted layers, which is challenging to achieve in other small-angle twisted devices, is a unique feature of our system. To quantify this, we calculate the probability that a moiré carrier is being simultaneously shared between both outer layers $h_{s,\boldsymbol{k}}=\sum_{\lambda=A,B}|\Psi_s^{t\lambda}(\boldsymbol{k})\Psi_s^{b\lambda}(\boldsymbol{k})|$, where $\Psi_s^{t/b\lambda}(\boldsymbol{k})$ denotes the wavefunctions at $\boldsymbol{k}$ for band $s$ of the electrons in the top or bottom layers and sublattice $\lambda$. When $h_{s,\boldsymbol{k}}$ is increased, carriers can efficiently tunnel between the outer layers, and in the opposite limit, $h_{s,\boldsymbol{k}}=0$, the tunneling is suppressed, allowing a layer-polarized regime. By integrating $h_{s,\boldsymbol{k}}$ over all states near the Fermi level we obtain a measure of total hybridisation $h$, as in Supplementary Equation (22) (Supplementary Note 5). However, we note that although the hybridization $h$ may appear analogous to the layer polarization frequently discussed in the literature, the two quantities can be clearly distinguished. Layer polarization simply describes the fraction of the total carrier density localized in each layer, whereas $h$ also accounts for the distributions of the corresponding wavefunctions. As a result, the same layer polarization state can correspond to several different layer hybridizations. Consequently, the two quantities exhibit clearly different behaviour in our system, as shown in Fig.~\ref{fig2}f and Extended Data Fig.~\ref{si_theory}c.
    
    The map of hybridization $h$ in Fig.~\ref{fig2}f closely follows the inverse DoS in Fig.~\ref{fig2}d, but also reveals a distinct belt of enhanced hybridization around $|D|/\varepsilon_0 \le 0.13\,$V/nm, coinciding with the region of increased $R_{\text{xx}}$ in Figs.~\ref{fig2}a,d. At larger $D$ and $n$, the hybridization is strongly suppressed at the corners of this map, and the bands become layer-polarized, consistent with the single-layer Landau quantization observed at the map corners in Fig.~\ref{fig2}a.
    
    Brown-Zak oscillations further demonstrate the tunability of interlayer hybridization. At zero $D$ (Fig.~\ref{fig3}a), strong oscillations appear across the full density range, reflecting strong hybridization. At finite displacement fields, $D/\varepsilon_0=0.3$ and $0.65\,$V/nm (Fig.~\ref{fig3}b,c), they become confined between the conductivity peaks corresponding to the stripe-like features of the double-gate map. This is consistent with the calculated hybridization map (Fig.~\ref{fig2}f), which predicts stronger hybridization within the horizontal belt and between the stripes, and decoupling at higher $n$, leading to the suppression of moiré effects. We also note that the interlayer hybridization generally decreases with increasing displacement field (inset of Fig.~\ref{fig3}b), which explains why Brown-Zak oscillations become less pronounced at higher $D$. In Fig.~\ref{fig3}c, one can still resolve the suppressed lines corresponding to even-denominator flux fillings 1/6, 1/4, and 1/2, which were also more pronounced at zero $D$, whereas the lines corresponding to odd-denominator flux fillings are less pronounced in this device and are more quickly suppressed with increasing $D$. Although the overall suppression of Brown–Zak oscillations with increasing $D$ is consistent with the calculated reduction in hybridization, their detailed visibility also depends on the local flatness of the magnetic Bloch bands~\cite{krishna2018high}, which may further influence their detailed behavior and requires separate analysis. We also note that even though the hybridization is reduced at high $D$, the layers remain coupled, as evidenced by the magnetic focusing measurements presented in Supplementary Note 6.

    \subsection*{Suppression of twist-angle disorder in TBG with a spacer}\vspace{-1.5em}

    We further find that the spacer stabilizes rigid moiré superlattices at marginal twist angles of approximately $0.1^{\circ}$ by suppressing atomic relaxation of the electronically active layers. In particular, our theoretical model, which does not include lattice-reconstruction effects, shows nearly ideal agreement with the measurements presented in Figs.~\ref{fig2},~\ref{fig3}. This is consistent with previous studies showing that atomic relaxation is strongly suppressed at large twist angles \cite{ezzi2024analytical}. In our devices, adjacent layers are typically misaligned by $20^{\circ}$--$30^{\circ}$, thereby minimizing reconstruction at their interfaces. Consequently, although the top and bottom graphene layers have a marginal relative twist angle $\sim 0.1^{\circ}$, the mechanical coupling transmitted through the spacer is not expected to induce appreciable reconstruction. We note that this situation differs from alternating magic-angle twisted trilayer graphene (MATTG), in which the relatively small twist angles of approximately $1.5^{\circ}$ between adjacent layers permit sufficiently strong mechanical coupling to promote the AA-stacking of the top and bottom layers \cite{turkel2022orderly}.

    Additionally, our devices systematically exhibit highly uniform twist angles, with variations of $\Delta\theta\lesssim0.02^{\circ}$ over distances of  $\sim 10\,\mu\mathrm{m}$. Measurements from CVD-grown device D3, shown in Fig.~\ref{fig4}a,b, provide a representative example. Brown--Zak oscillations measured in different segments of the device distributed over approximately $10\,\mu\mathrm{m}$ show only small variations in their magnetic-field positions and, consequently, in the extracted twist angles. The twist-angle disorder within each segment is also smaller than $0.02^{\circ}$. The presence of the local disorder would broaden the peaks corresponding to a given rational flux fraction. However, the characteristic half-widths of these peaks (e.g., around flux one), even though they already start to broaden due to temperature increase, are comparable to or smaller than the differences between their positions in different segments, placing an upper bound on the local twist-angle disorder.

    Similar behaviour is also observed in the other devices. For example, CVD-grown device D1, shown in Extended Data Fig.~\ref{CVD_data}, exhibits comparably uniform Brown--Zak oscillations in different device regions. Manually stacked devices D1 and D2, which were fabricated from the same heterostructure but are separated by approximately $10\,\mu\mathrm{m}$, also exhibit a twist-angle variation of only $\Delta\theta\lesssim0.02^{\circ}$. Overall, our measurements indicate that twist-angle disorder in twist-with-a-spacer devices is smaller than the variations of approximately $0.1^{\circ}$ or more typically observed over comparable distances in directly contacting TBG devices \cite{uri2020mapping, kazmierczak2021strain}.

    We point out, however, that both the twist-angle disorder and the suppression of atomic reconstruction in our devices are inferred indirectly from electronic-transport measurements. Direct visualization of the moiré periodicity and atomic structure using scanning-probe or electron-microscopy techniques would therefore be valuable for further characterization of these superlattices.
     
    Finally, we also find that the Brown-Zak oscillations in TBG-with-a-spacer are only weakly sensitive to temperature, Fig.~\ref{fig4}c, and can even persist up to room temperature, Extended Data Fig.~\ref{si3}d,e. Such room-temperature persistence has previously been observed in high-quality graphene/hBN superlattices \cite{krishna2017high,krishna2018high} but, to the best of our knowledge, has not been reported in other moiré systems.

    \subsection*{Conclusions and outlook} \vspace{-1.5em}  
    
    Overall, our results demonstrate that twisted graphene layers separated by a monolayer spacer retain sufficient interlayer coupling to form well-defined moiré minibands with tunable hybridization. This realizes a qualitatively distinct regime of stacked two-dimensional electron systems in which the electronically active layers are not in direct contact, yet remain coupled by interlayer tunnelling strong enough to substantially reconstruct their band structure. At the same time, the spacer suppresses atomic relaxation of the active layers, stabilizing rigid moiré superlattices with extremely small and highly uniform twist angles down to $\theta \sim 0.1^\circ$.

    This structural stability offers a route towards overcoming a central challenge in the scalable growth of small-angle moiré heterostructures. Although previously manually assembled twisted structures have revealed a broad range of unconventional phenomena with potential relevance to future electronic devices, their direct and controllable scalable growth remains challenging. During growth, elevated temperatures drive graphene layers separated by a small twist angle to relax towards the energetically favourable Bernal-stacked configuration, making small-angle grown structures relatively unstable \cite{cho2020specific,sun2021hetero,liu2022designed,liu2025oxygen}. Consequently, reproducible growth of twisted graphene heterostructures has so far been largely restricted to twist angles above $\approx3$-$5^\circ$. Although a small fraction of grown crystals in these works exhibited small twist angles, electronic transport measurements probing moiré phenomena in such crystals have yet to be demonstrated.
    
    Moiré coupling through a spacer provides a solution to this problem. In our CVD-grown structures, a strongly misaligned middle graphene layer prevents two outer layers from collapsing into Bernal stacking. Nevertheless, tunnelling through this spacer remains sufficiently strong to produce highly tunable moiré minibands. Importantly, the resulting CVD-grown devices exhibit electronic quality comparable to that of devices fabricated from exfoliated and manually stacked graphene layers.
    
    Although our growth process should, in principle, produce perfectly aligned outer graphene layers, all Hall bars studied here exhibit small twist angles of $0.17$-$0.22^\circ$ (Supplementary Table 1), whose origin requires further investigation. These twist angles may be introduced during transfer from the Cu growth substrate to SiO$_2$ or during subsequent heterostructure assembly. However, we cannot exclude the possibility that a small spontaneous misalignment develops during CVD growth. Further optimization of the growth and transfer procedures could potentially enable CVD-grown devices with twist angles approaching the theoretically predicted magic angle of this system, $\approx 0.04^\circ$. This would open a route towards studying correlated phenomena in scalable CVD-grown structures, which is currently challenging to achieve through manual stacking in the introduced twist-with-a-spacer geometry. We note, however, that such an extremely small twist angle corresponds to a moiré periodicity of approximately $350~\mathrm{nm}$, implying that the resulting devices must have sufficiently large lateral dimensions to observe the relevant moiré phenomena in this limit.
    
    Beyond scalability, the twist-with-a-spacer architecture provides a versatile platform for exploring and engineering moiré band structures. We demonstrate that the reduced interlayer coupling allows the band structure and layer hybridization to be efficiently controlled using moderate displacement fields, which is difficult to achieve in conventional small-angle TBG. By combining transport measurements with theoretical calculations, we disentangle the respective roles of interlayer hybridization and superlattice periodicity in determining the resulting miniband structure (Extended Data Fig.~\ref{si_theory}). This separation provides a useful design tool for engineering moiré electronic systems in which periodicity and coupling can be controlled independently.
    
    
\subsection*{Data availability}  \vspace{-1.5em}   
All the data shown in this work are available from the corresponding authors upon a reasonable request. 
\subsection*{Competing interests}  \vspace{-1.5em}   
The authors declare no competing interests.

\subsection*{Acknowledgments}  \vspace{-1.5em}   
This work has been supported by the National Research Foundation, Singapore, under its NRF Fellowship (NRFF) (NRF-NRFF16-2024-0011), and by
A*STAR under its RIE2025 Manufacturing, Trade and Connectivity (MTC)
Young Individual Research Grant (M23M7c0126) awarded to A.I.B.; 
K.W. and T.T. acknowledge support from the JSPS KAKENHI (Grant Numbers 21H05233 and 23H02052), the CREST (JPMJCR24A5), JST and World
Premier International Research Center Initiative (WPI), MEXT, Japan. 
K.S.N. acknowledges support from the Royal Society
(UK, grant no. RSRP/R/190000). This project is supported by the National
Research Foundation, Singapore under its AI Singapore Programme (AISG
Award No: AISG3-RP-2022-028). 
B.F. thanks Quantum funding in EPSRC DTP grant EP/W524347/1.
V.F. acknowledges support from  EPSRC Grants EP/S030719/1 and EP/V007033/1, and the Lloyd Register Foundation Nanotechnology Grant.
The collaboration aspect of this project was supported by the International Science Partnerships Fund (UK). We also thank Eli Rotenberg, Aaron Bostwick, and Chris Jozwiak for their support with the ARPES measurements performed at the Advanced Light Source, Berkeley (CA), USA, which provided complementary characterization of the CVD-grown graphene structures.

\bibliographystyle{naturemag}
\bibliography{ref}

\newpage
    \begin{ExtFigure}[t]
	\includegraphics[width=\linewidth]{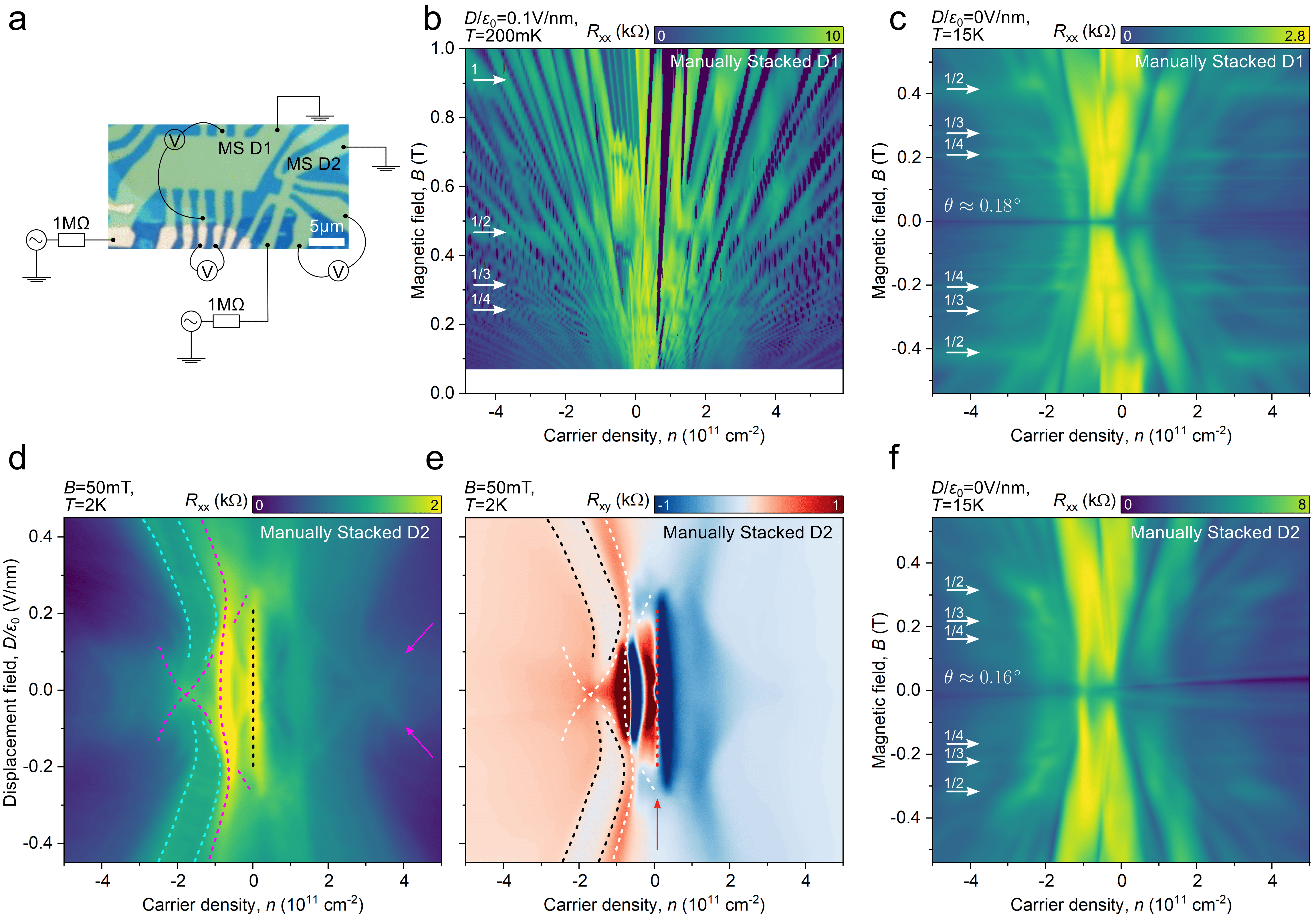}
	\caption{\textbf{Additional data from Manually Stacked Devices 1 and 2 and demonstration of twist-angle homogeneity over distances of $\approx 10\,\mu\mathrm{m}$.}
    \textbf{a}, Optical image of the Manually Stacked Devices 1 and 2 fabricated from the same TBG with a monolayer graphene spacer stack. The measurement configurations indicated in the optical image correspond to those used to acquire the fan diagrams displayed in panels (c) and (f). 
    \textbf{b}, Fan diagram measured in Manually Stacked D1 under an applied displacement field of $D=0.1$V/nm at 200mK. 
    \textbf{c}, Fan diagram measured in Manually Stacked D1, for $D=0$V/nm at an elevated temperature of 15K. 
    \textbf{d, e}, Longitudinal and Hall resistance measured in Manually Stacked D2 at magnetic field 50 mT as a function of $n$ and $D$. In panel (d), cyan dashed lines indicate examples of features originating from pure layer hybridization, while magenta dashed lines and arrows mark examples arising from superlattice-enabled umklapp. The dashed black line indicates a resistance peak that is not consistent with a single-particle picture. Similarly, in panel (e), black dashed lines indicate examples of features originating from layer hybridization, while the white dashed line corresponds to features arising from the superlattice potential. The dashed red line and arrow highlight the feature identified in panel d that is not consistent with single-particle modeling. 
    \textbf{f}, Fan diagram measured in Manually Stacked D2, for $D=0$V/nm at an elevated temperature of 15K. The suppressed contrast of Brown-Zak oscillations in this device in comparison with Manually Stacked D1 likely originates from the characteristic dimensions of the devices, as Manually Stacked D1 is almost four times larger than Manually stacked D2. At these small twist angles, the large moiré wavelength of approximately $\sim80$nm becomes non-negligible relative to the Hall-bar dimensions, potentially reducing the visibility of moiré-related transport phenomena.}
	\label{si4}
\end{ExtFigure}

\newpage
    \begin{ExtFigure}[t]
    \includegraphics[width=\linewidth]{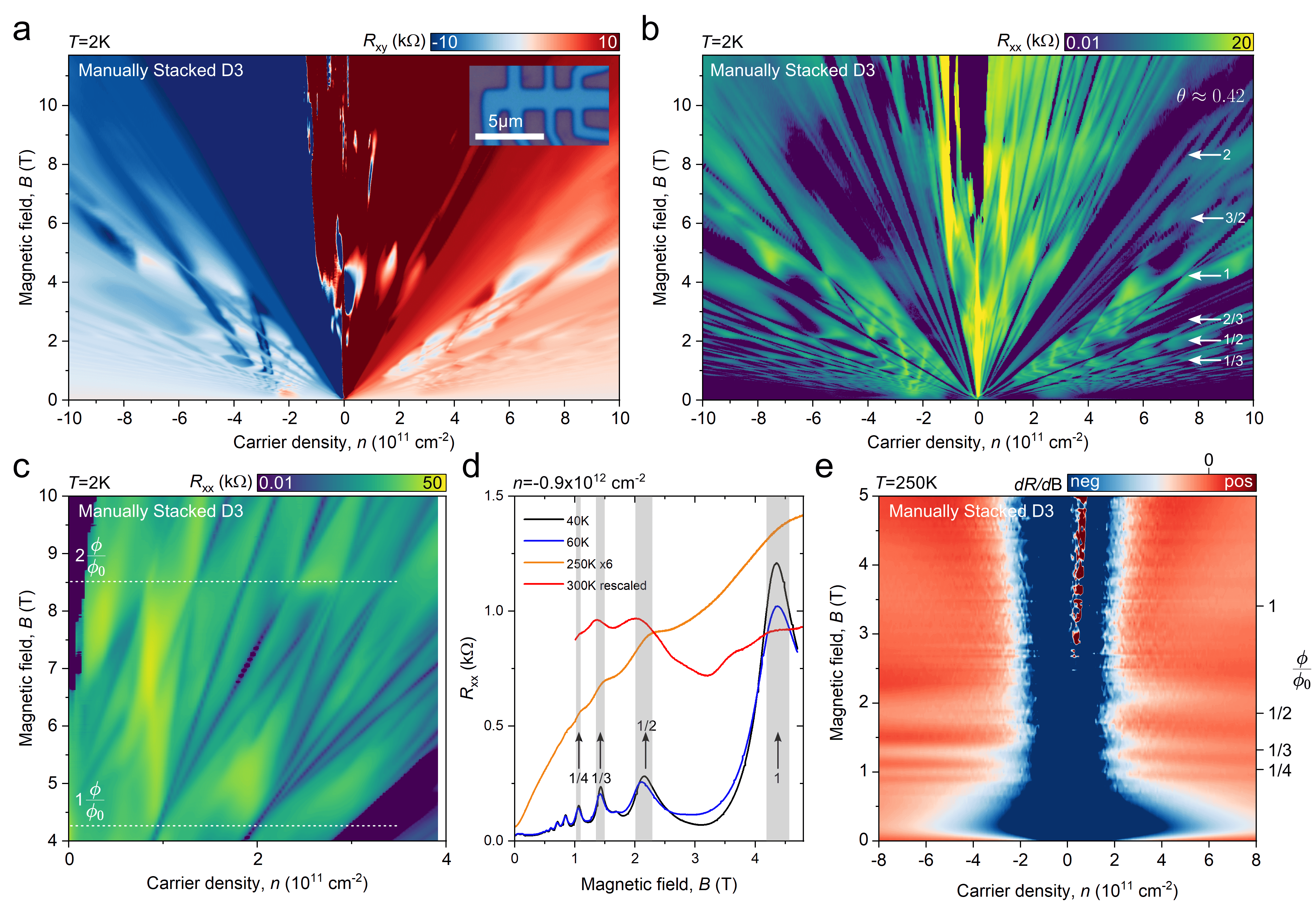}
    \caption{\textbf{Characterization of Manually Stacked D3, and persistence of Brown-Zak oscillations upto room temperature}. 
    \textbf{a, b}, $R_{\text{xy}}$ and $R_{\text{xx}}$ maps as a function of $n$ and $B$. The inset in panel (a) shows an optical image of Manually Stacked D3. Horizontal arrows in panel (b) mark the positions of Brown-Zak oscillations, used to determine the twist angle $\theta \approx 0.42^{\circ}$.
    \textbf{c}, Zoomed-in view of the Landau fan shown in panel (b). Dashed lines indicate one- and two-flux-quanta fillings per superlattice unit cell. 
    \textbf{d}, Brown–Zak oscillations measured as a function of magnetic field at different temperatures. To enhance the visibility of features at high temperature, we multiplied the data at 250~K by 6, and at 300~K we subtracted a linear function from the data and then multiplied it by 10. 
    \textbf{e}, Derivative of resistance as a function of magnetic field and carrier density measured at 250~K, showing the persistence of Brown-Zak oscillations at elevated temperatures.}
\label{si3}
\end{ExtFigure}

\newpage
    \begin{ExtFigure}[t]
    \includegraphics[width=\linewidth]{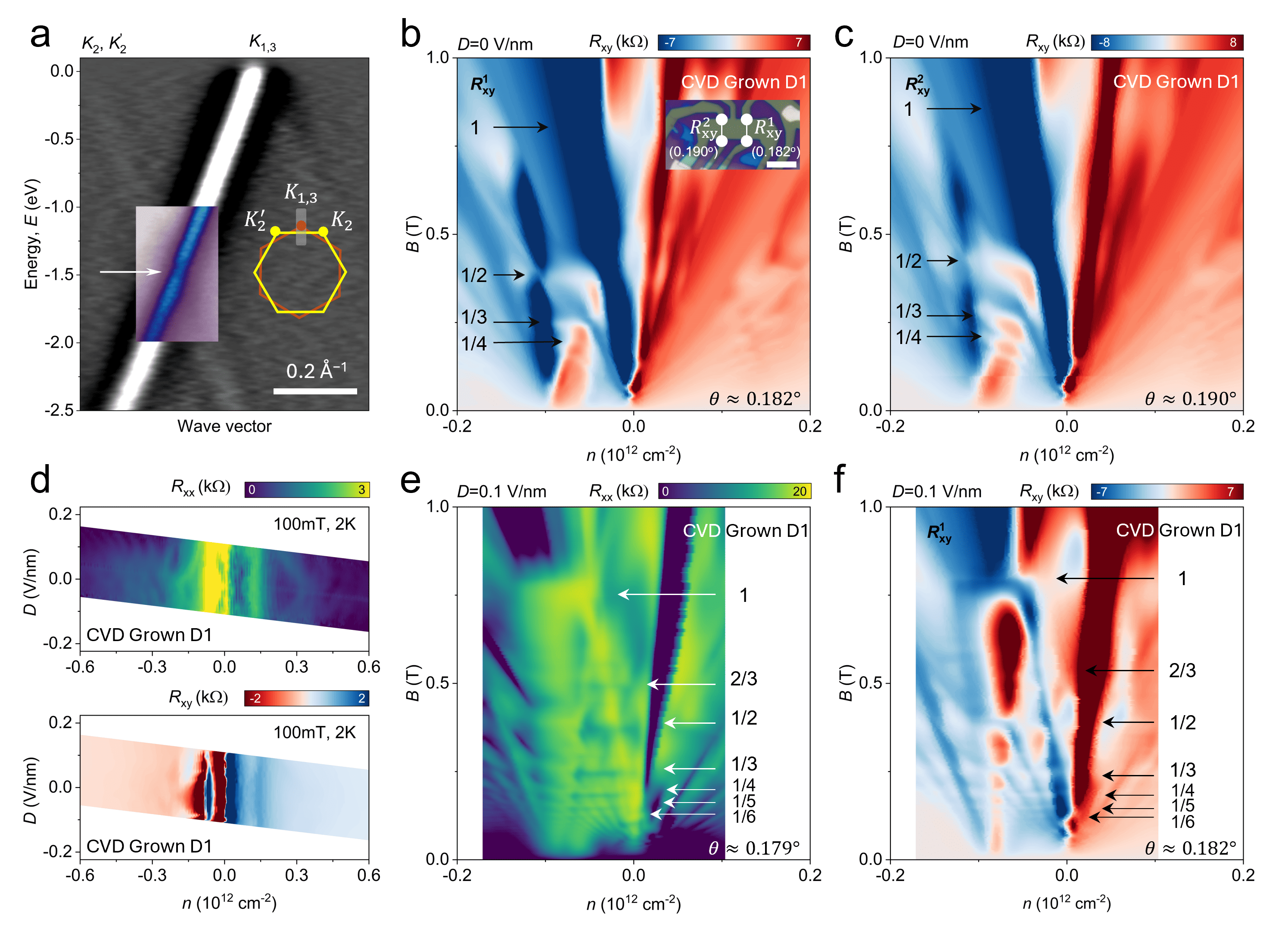}
    \caption{\textbf{Additional data from CVD-grown Device 1.} 
    \textbf{a}, Nano-ARPES measurements of CVD-grown TBG with a graphene spacer. Inset, full Brillouin zones of the TBG with spacer, showing the Dirac points of the top and bottom graphene layers ($K_{1,3}$) and the middle graphene layer ($K_2$). The measurement was performed along the grey region passing through $K_{1,3}$. An ARPES inset with enhanced colour contrast is included to improve the visibility of the gap. The white arrow indicates an avoided crossing at approximately $-1.5$ eV. Characteristically, only one side of the top-layer Dirac cone is visible, due to the sublattice composition of Dirac electrons \cite{mucha2008characterization, hamer2022moire}. 
    \textbf{b, c}, Hall-effect maps measured at constant $D = 0$ V nm$^{-1}$ in two different regions of the device, as indicated in the optical image (inset in (b)). The scale bar for the device image is $5\,\mu\mathrm{m}$. Horizontal arrows indicate the positions of Brown-Zak oscillations, which were used to determine the local twist angle. 
    \textbf{d}, $R_{\mathrm{xx}}$ and $R_{\mathrm{xy}}$ maps as a function of $n$ and $D$, measured at a magnetic field of 100 mT. Leakage through the top-gate dielectric occurred at relatively low gate voltages in this device, limiting the accessible range of $D$. 
    \textbf{e, f}, $R_{\mathrm{xx}}$ and $R_{\mathrm{xy}}$ maps as a function of $n$ and $B$, measured at $D = 0.1$ V nm$^{-1}$. Horizontal arrows indicate the positions of Brown-Zak oscillations, which were used to determine the local twist angle. All measurements were performed at 2 K.}
    
    \label{CVD_data}
\end{ExtFigure}

\newpage
\begin{ExtFigure}[t]
	\includegraphics[width=\linewidth]{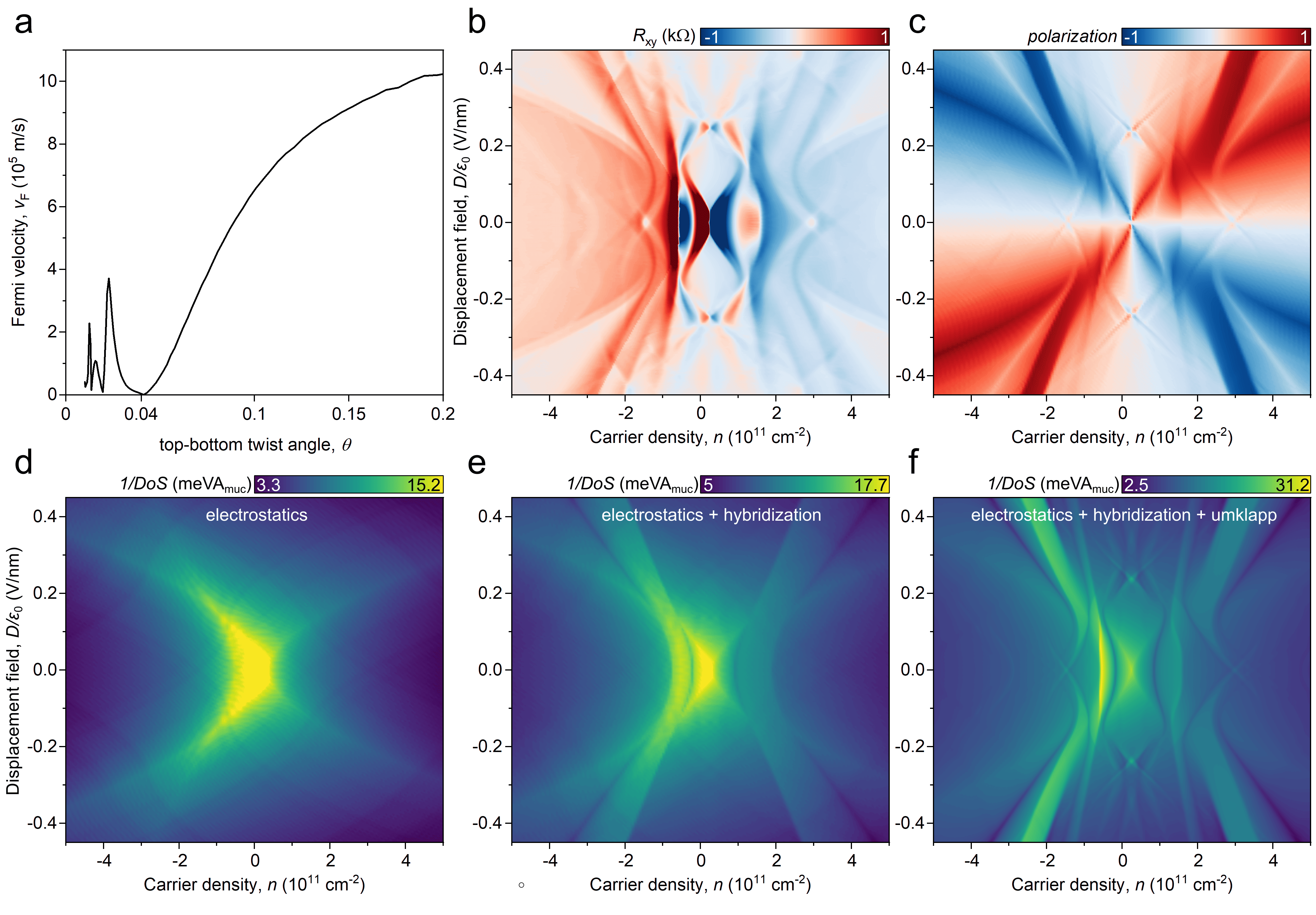}
    \caption{\textbf{Disentangling the roles of interlayer hybridization and superlattice periodicity in miniband formation}.
    \textbf{a}, Magnitude of the velocity of the first/lowest conduction band in the effective model for the outer layers at the $\kappa$ point as a function of twist angle $\theta$. The first magic angle, where the velocity is quenched, occurs at $0.04^\circ$. 
    \textbf{b}, Theoretically calculated Hall resistance $R_{xy}$ as a function of $n$ and $D$, calculated from the sum of conductivities of the outer layers and spacer layer as in Supplementary equation (20) with $\tau_\text{sp}/\tau_\text{m}=0.3$. See Supplementary Note 4 for further details.
    \textbf{c}, A polarization map, where red (blue) indicates the relative weight of the carrier density in the top (bottom) layer.
    \textbf{d}-\textbf{f}, Inverse DoS maps calculated for the TBG with a monolayer graphene spacer, with different contributions included according to the labels in the panels.}
	\label{si_theory}
\end{ExtFigure}

\end{document}